\documentclass[amssymb,prl,twocolumn,floats]{revtex4}
\usepackage{bm}
\usepackage{graphicx}
\begin{document}
\preprint{September 14, 2001}
\title{Ferromagnetism of magnetic semiconductors---Zhang-Rice limit}
\author{T. Dietl$^{(a)}$,
F. Matsukura$^{(a,b)}$, and H. Ohno$^{(b)}$}
\affiliation{$^{(a)}$Institute of Physics, Polish Academy of
Sciences, al. Lotnik\'{o}w 32/46, PL-00-668 Warszawa, Poland\\
$^{(b)}$Laboratory for Electronic Intelligent Systems, Institute of
Electrical Communication, Tohoku University, Katahira 2-1-1, Sendai
980-8577, Japan}
\date{\today}
\begin{abstract}
It is suggested that p-d hybridization contributes significantly to
the hole binding energy $E_b$ of Mn acceptors in III-V compounds,
leading in an extreme case to the formation of Zhang-Rice-like
small magnetic polarons. The model explains both strong increase of
$E_b$ and evolution of Mn spin-resonance spectrum with the
magnitude of valence-band offsets. For such a structure of Mn
impurity in III-V materials, possible models accounting for the
recently determined Curie temperature of about 940~K in a
compensated Ga$_{0.91}$Mn$_{0.09}$N are discussed.
\end{abstract}

\maketitle

Recent theoretical suggestions \cite{Diet00,Sato01} that
ferromagnetic ordering may persist above room temperature in
(Ga,Mn)N and related materials have triggered considerable
fabrication efforts \cite{Akin00,Zaja01,Over01}, which have
culminated in the determination of Curie temperature
$T_{{\mbox{\small C}}} \approx 940$~K for a film of a compensated
Ga$_{0.91}$Mn$_{0.09}$N frown by molecular-beam epitaxy (MBE)
\cite{Sono01}. In order to put this result in an appropriate
perspective we note that the magnetic ion density is about twenty
times smaller in Ga$_{0.91}$Mn$_{0.09}$N comparing to Co, which has
the highest known value of $T_{{\mbox{\small C}}} = 1400$~K. This
adds a new dimension to capabilities of GaN-based structures, whose
importance in photonics and high-power electronics has already been
proven. While further experimental studies of this new
ferromagnetic system will certainly lead to developments that are
unforeseen today, it is already important to model the physical
environment, in which the ferromagnetism is observed as well as to
indicate parameters that are controllable experimentally, and which
account for the magnitudes of $T_{{\mbox{\small C}}}$ and remanent
magnetization.

In this work, we demonstrate that the high $T_{{\mbox{\small C}}}$
value quoted above is consistent with the expectations of the
mean-field Zener model of the carrier-mediated ferromagnetism,
which was elaborated earlier by us and co-workers
\cite{Diet00,Diet01}. We then examine various assumptions and
approximations of the model exploring recent developments in the
field of magnetic semiconductors
\cite{Sato01,Schl00,Bhat01,Chat01,Wolo01,Blin01}. In particular, by
a detail tracing of properties of Mn impurity across the ensemble
of III-V semiconductors we suggest that Zhang-Rice (ZR) small
magnetic polarons are formed in GaN:Mn. We then propose that
delocalization of these polarons may drive the ferromagnetic
transition in p-(Ga,Mn)N. We also note that if the d$^5$/d$^6$
level resides in the gap and the concentration of donors is greater
than that of Mn, double exchange involving the upper Hubbard d-band
may account for ferromagnetism of n-(Ga,Mn)N.

Figure 1 presents $T_{{\mbox{\small C}}}$ of wurzite
Ga$_{1-x}$Mn$_x$N calculated according to the model
\cite{Diet00,Diet01} and for material parameters discussed in
detail previously \cite{Diet01}. The data are shown for
magnetization perpendicular to the $c$ axis of the wurzite
structure, which is found to be the easy plane in (Ga,Mn)N for the
displayed values of the Mn and hole concentrations. As seen,
$T_{{\mbox{\small C}}}$ of 940~K is expected for the hole
concentrations about four times smaller than the Mn concentration
$x$ in Ga$_{0.91}$Mn$_{0.09}$N. Such a value of $T_{{\mbox{\small
C}}}$ is consistent with that deduced from magnetization
measurements at 0.1 T between 300 and 750~K \cite{Sono01}.

We begin the discussion of various features of theory
\cite{Diet00,Diet01} by recalling that according to the mean-field
Zener model \cite{Zene50}, the properties of the system consisting
of localized spins and itinerant carriers can be obtained by
minimizing the total free energy with respect to the spin
magnetization $M$. It is assumed \cite{Diet00,Diet01} that the Mn
ions substitute the cations, and are in the 3d$^5$ $S=5/2$
configuration, whereas the holes reside in the valence band, which
is described by the disorder-free $k\cdot p$ Kohn-Luttinger theory
for tetrahedrally coordinated semiconductors. The p-d exchange
between the two subsystems, the Kondo interaction, is incorporated
to the $k\cdot p$ scheme within the molecular-field and virtual
crystal approximations, whereas the hole-hole exchange interaction
is treated in terms of Landau's Fermi liquid theory. The hole free
energy, not ground state energy, is adopted for the evaluation of
$T_{{\mbox{\small C}}}$ \cite{Diet00,Diet01}, an ingredient
significant quantitatively in the high temperature regime, such as
that covered by the data in Fig.~1.

Figure 2 shows the energetic position of the Mn impurity level in
III-V compounds, as evaluated by various authors from measurements
of optical spectra and activation energy of conductivity. {\em A
priori}, the Mn atom, when substituting a trivalent metal, may
assume either of two configurations: (i) d$^4$ or (ii) d$^5$ plus a
weakly bound hole (d$^5$+h $\equiv$ A$^0$). Accordingly, the
experimentally determined energies correspond to either d$^4$/d$^5$
or A$^0$/A$^-$ levels. It appears to be a general consensus that
the Mn acts as an effective mass acceptor (d$^5$+h) in the case of
antimonides and arsenides. Such a view is supported by a relatively
small Mn concentrations corresponding to the insulator-to-metal
transition, which according to the Mott criterion $n^{1/3}a_B =
0.26$, points to a relatively large extension of the effective Bohr
radius $a_B$. Moreover, the ESR studies of GaAs:Mn reveal, in
addition to the well known spectrum of Mn d$^5$ with the Land\'e
factor $g_S=2.0$, two additional lines corresponding to $g_a=2.8$
and $g_b=6$ \cite{Schn87,Mast88,Szcz99}, which can be described
quantitatively within the $k\cdot p$ scheme for the occupied
acceptor \cite{Schn87,Mast88}. Here, the presence of a strong
antiferromagnetic p-d exchange interaction between the bound hole
and the Mn d-electrons has to be assumed, so that the total
momentum of the complex is $J=1$. In agreement with the model, the
additional ESR lines, in contrast to the $g=2.0$ resonance, are
visible only in a narrow range of the Mn concentration
\cite{Szcz99}, which should be greater than the concentration of
compensating donors, and smaller than that at which acceptor wave
functions start to overlap and to merge with the valence band. The
antiferromagnetic coupling within the d$^5$+h complex is seen in a
number of experiments, and has been employed to evaluate the p-d
exchange integral $\beta N_0 \approx - 1$~eV \cite{Bhat00} in
GaAs:Mn, the value in agreement with that determined from interband
magnetoabsorption in (Ga,Mn)As \cite{Szcz01}.

\begin{figure}
\includegraphics*[width=90mm]{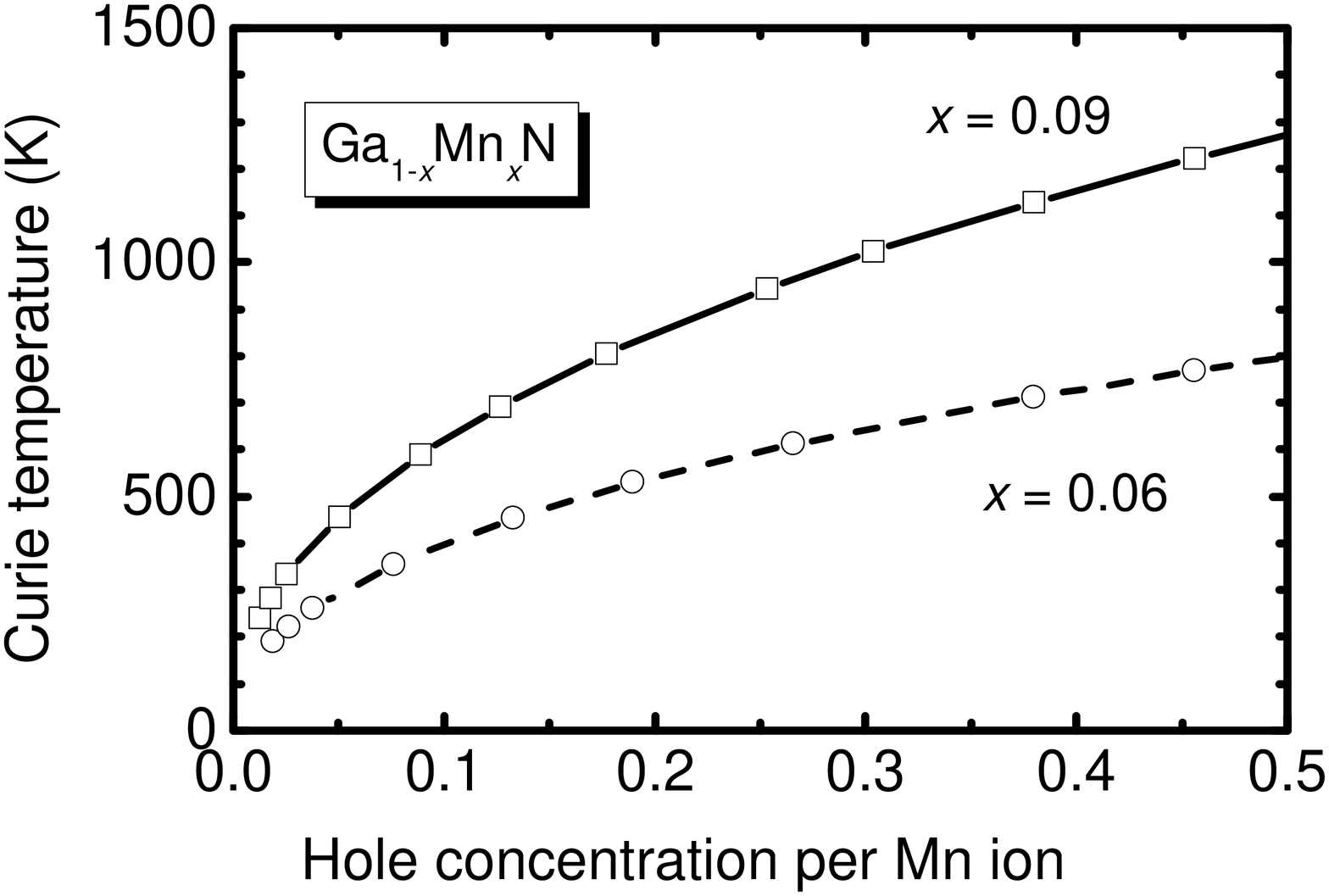}
\caption[]{Curie temperature of wurzite Ga$_{1-x}$Mn$_x$N according
to the mean-field Zener model \cite{Diet00,Diet01} calculated as a
function of the hole concentration per Mn ion, $x_h = p/xN_o$.}
\label{fig:T_C}
\end{figure}

Importantly, the above scenario is corroborated by results of
photoemission \cite{Okab98,Okab01} and x-ray magnetic circular
dichroism (MCD) studies \cite{Ohld00,Ueda01} in metallic or nearly
metallic Ga$_{1-x}$Mn$_{x}$As. The latter point to the d$^5$ Mn
configuration. The former are not only consistent with such a
configuration but also lead to the value of $\beta N_0$ similar to
that quoted above, $\beta N_0 \approx - 1.2$~eV \cite{Okab98}.
Furthermore, the photoemission reveals the presence of two features
in the density of states brought about by the Mn constituent: the
original Mn 3d$^5$ states located around 4.5~eV below the Fermi
energy $E_F$, and new states merging with the valence band in the
vicinity of $E_F$ \cite{Okab01}. These new states correspond to
acceptors, as discussed above. They are derived from the valence
band by the Coulomb field as well as by a local Mn potential that
leads to a chemical shift in the standard impurity language, or to
a valence band offset in the alloy nomenclature.  The fact that
transition metal impurities may produce both resonant and Coulomb
states was also noted in the case of CdSe:Sc \cite{Glod94}.

In contrast to antimonides and arsenides, the situation is much
more intricate in the case of phosphides and nitrides. Here, ESR
measurements reveal the presence of a line with $g=2.0$ only
\cite{Zaja01,Krei96,Kiri82,Dawe83,Mast81}, which is thus assigned
to d$^5$ (A$^-$) centers \cite{Zaja01,Krei96,Kiri82,Dawe83}.
Moreover, according to a detail study carried out for a compensated
n-type GaP:Mn \cite{Krei96}, the ESR amplitude diminishes under
illumination and, simultaneously, new lines appear, a part of which
exhibit anisotropy consistent with the d$^4$ configuration. This,
together with the apparent lack of evidence for d$^5$+h states,
even in p-type materials, appears to imply that Mn in the ground
state possesses the d$^4$, not A$^0$, electron configuration
\cite{Krei96}. This would mean that the Mn energy in Fig.~2 for GaP
\cite{Krei96} and, therefore, for GaN \cite{Wolo01,Blin01} (where
the valence band is lower than in GaP) corresponds to the
d$^4$/d$^5$, not A$^0$/A$^-$ level. Such a view seems to be
supported by the {\em ab initio} computation within the local spin
density approximation (LSDA), which points to the presence of the
d-states in the gap of (Ga,Mn)N \cite{Sato01}. In this situation,
the spin-spin interaction would be driven by a double exchange
mechanism involving hopping of d-electrons \cite{Sato01,Zene51}, as
in the case of colossal magnetoresistance manganites, making the
modeling leading to the results presented in Fig.~1 invalid.

\begin{figure}
\includegraphics*[width=90mm]{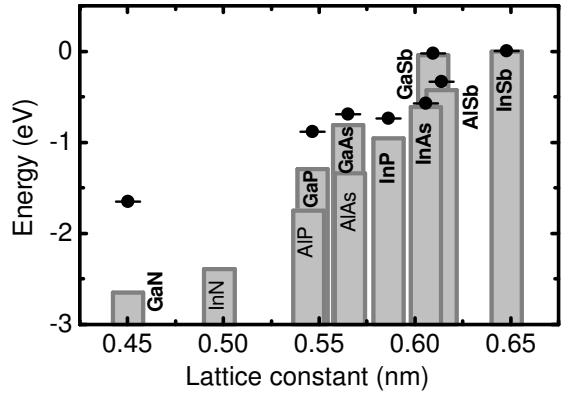}
\caption[]{Experimental energies of Mn levels in the gap of III-V
compounds according to
Refs.~\cite{Wolo01,Krei96,Linn97,Lamb85,Parr94,Mare94,Hofm93,Obuh97})
in respect to valence-band edges, whose relative positions are
taken from Ref.~\cite{Vurg01}.}
 \label{fig:III-V}
\end{figure}

However, there is a chain of arguments which calls the above
interpretation into question. First, guided by photoemission
results for II-VI compounds \cite{Mizo93} one expects that the
energy of the d$^4$/d$^5$ level will vary little between arsenides
and nitrides. This implies that this level should reside in the
valence band of GaN despite the 1.8~eV valence band offset between
GaN and GaAs. The resulting contradiction with the LSDA findings
can be removed by noting that in the case of strongly correlated 3d
electrons, a semi-empirical LSDA+U approach is necessary to
reconcile the computed and photoemission positions of states
derived from the Mn 3d shell in (Ga,Mn)As \cite{Okab01,Park00}.

Second, it is known that the p-d hybridization, in addition to
producing the exchange interaction, can contribute to the hole
binding energy $E_b$ \cite{Zhan88,Beno92}. We take the hole wave
function as a coherent superposition of p-stated of anions adjacent
to Mn \cite{Zhan88}. Assuming the p-orbitals to be directed towards
the ion Mn we find that the $T_2$ state has 30\% lower energy than
that corresponding to the mutually parallel p-orbitals. This shows
that not only Kohn-Luttinger amplitudes from the $\Gamma$ point of
the Brillouin zone are involved.  In order to evaluated $E_b$,  a
square-well spherical potential $U(r) = U\Theta(b-r)$ is assumed
\cite{Beno92}, whose depth $U$ is determined by the p-d
hybridization taking into account the above mentioned arrangement
of the p-orbitals,
\begin{equation}
U=\frac{5}{8\pi}\frac{\beta
N_0}{1.04}\left(1-\frac{\Delta_{eff}}{U_{eff}}\right)
\left(\frac{a_o}{b}\right)^3.
\end{equation}
Here, the values of $\beta N_0$ are taken from Ref.~\cite{Diet01};
$\Delta_{eff}$ is the distance of d$^4$/d$^5$ level to the top of
the valence band, which is evaluated to be 2.7~eV in (Ga,Mn)As
\cite{Okab98}, and is assumed to be reduced in other compounds by
the corresponding valence band offsets (Fig.~2), and $U_{eff}=
7$~eV is the correlation energy of the 3d electrons
\cite{Okab98,Mizo93}. Finally, $b/a_o$ is the well radius in the
units of the lattice constant, and should lie in-between the
cation-anion and cation-cation distance, $\sqrt{3}a_o/4 <b<
a_o/\sqrt{2}$. It turns out that in the case of GaN:Mn the hole is
bound by Mn, even in the absence of the Coulomb potential, $E_b
=1.0$~eV for $b = 0.46a_o$. This demonstrates rather convincingly
that a large part of $E_b$ originates indeed from the p-d
interaction, indicating that the Zhang-Rice (ZR) limit
\cite{Zhan88,Beno92} is reached in these systems. The formation of
ZR small magnetic polarons invoked here has a number of important
consequences which are now discussed.

We begin by arguing that when the ZR polarons are formed, the d$^5$
and d$^5$+h states acquire virtually the same Land\'e factor
$g=2.0$. Quite generally, the role of the intra-atomic spin-orbit
splitting $\Delta_o$ diminishes with the carrier kinetic energy,
and thus with $E_b$. An upper limit of orbital-momentum corrections
to the Mn Land\'e factor, $\Delta g = g - g_S$ can thus be
evaluated by considering non-magnetic acceptors, for which $E_b$ is
typically smaller than that of Mn. In the wurzite GaN, the orbital
momentum is quenched by the uniaxial crystal field, and the Land\'e
factors of GaN:Mg are close to two, $g_h^{\parallel} = 2.07$ and
$g_h^{\perp} = 1.99$ \cite{Palc98}. Hence, for the d$^5$+h state in
wz-GaN:Mn, where the total spin $S_p=2$, we obtain $\Delta g =
(g_S-g_h)/6 \approx -0.01$. In cubic materials, in turn, the total
momentum $J_p=1$ and $\Delta g=3/4$ for deep d$^5$+h states
\cite{Schn87,Mast88}, in agreement with the ESR findings for
GaAs:Mn \cite{Schn87,Mast88,Szcz99}. We note, however, that the
Jahn-Teller effect should be particularly strong in the case of ZR
polarons. The corresponding lowering of local symmetry will lead to
the quenching of orbital momentum, especially in compounds with a
relatively small value of $\Delta_o$. Hence, also in the cubic
nitrides and phosphites, $S_p=2$ and thus $\Delta g \approx 0$.
Interestingly, by exploiting the hole interaction with host nuclear
spins, it becomes possible to tell d$^5$ and d$^5$+h states despite
their similar Land\'e factors \cite{Mast81}. In particular, the
presence of the hole explains a surprising broadening of the Mn
hiperfine sextet already at relatively small concentrations of Mn
impurities, observed in InP:Mn \cite{Dawe83,Mast81,Sun92}, GaP:Mn
\cite{Kiri82}, and GaN:Mn \cite{Zaja01} as well the detection of P
and In ENDOR signals in InP:Mn \cite{Sun92}.

Another important consequence of the the ZR polaron formation is
the shift of the Mott critical concentration towards rather high
values. According to the known relation between $E_b$ and $a_B$
\cite{Bhat01}, the critical hole concentration is $p_c=4\times
10^{19}$ cm$^{-3}$ in (Ga,Mn)As and at least an order of magnitude
greater in (Ga,Mn)N, if no shallower acceptors are present.
According to the two-fluid model \cite{Diet00}, corroborated by
experimental results \cite{Ohld00,Oiwa97,Ferr01}, only a part of
spins orders ferromagnetically in the insulator phase. The Mn spins
are ordered in space regions, in which weakly localized holes
reside \cite{Diet00,Bhat01}. This can be the case of the
ferromagnetic Ga$_{0.91}$Mn$_{0.09}N$ sample \cite{Sono01}, whose
spontaneous magnetic moment constitutes only about 20\% of the
value expected for $x = 0.09$ and $S = 5/2$. Remarkably, the
disorder in Mn positions seems to shift $T_{{\mbox{\small C}}}$ to
higher values in such a case \cite{Bhat01}, presumably even above
those display in Fig.~1. On the other hand, corrections to
molecular- and mean-field approximations tend to lower
$T_{{\mbox{\small C}}}$ values \cite{Schl00,Chat01}. While our work
provides a background for detail studies of these effects, we
expect that the corresponding corrections will not be qualitatively
significant. Indeed, the short-range potential of Eq.~1 is still
relatively weak: our $|\beta|$ corresponds to $J_{pd} =
0.054$~eVnm$^3$ and $J$ = 0.5~eV of Refs.~\cite{Schl00,Chat01},
respectively.

It might appear from Fig.~1 that a further enhancement of
$T_{{\mbox{\small C}}}$ would be possible by increasing the hole
concentration $p$. Actually, however, if $p \rightarrow xN_0$ ({\em
i.e.}, $x_h\rightarrow 1$), the antiferromagnetic portion of the
carrier-mediated (RKKY) interaction will reduce $T_{{\mbox{\small
C}}}$ \cite{Diet00,Schl00,Chat01,Ferr01} and eventually drive the
system towards a spin-glass phase \cite{Egge95}. Thus, it seems
that the compensation by donors, despite introducing an additional
disorder, constitutes the important ingredient of ferromagnetism in
Mn-based III-V compounds \cite{Remark}.

Finally, we note that if indeed, as we advocate here, the
d$^4$/d$^5$ state resides in the valence band, the d$^5$/d$^6$
level may lie below the bottom of the conduction band. If this is
the case, and the concentration of donors is greater than that of
Mn, double exchange involving the d$^6$ electrons may account for
ferromagnetism of n-(Ga,Mn)N. We note in this context that because
of the anomalous (spin) Hall effect \cite{Omiy00}, a possible
coexistence of p-type and n-type regions in compensated
semiconductors \cite{Efro84}, and the presence of strong electric
fields due to lack of inversion symmetry, the evaluation of the
carrier type and concentration from transport measurements is by
no means straightforward in thin films of wz-(Ga,Mn)N.

In summary, according to the discussion presented in this paper,
while the Mn substitutional impurities act as shallow acceptors
characterized by the Bohr radius spacing over tens of {\AA} in
antimonides, in the case of lighter anions the hole becomes tightly
localized. In the extreme case, such as GaN:Mn, the Zhang-Rice-like
magnetic polarons are formed. At sufficiently high Mn
concentrations, the holes tend to delocalize and can then mediate
long-range correlation between the Mn ions.  In an overcompensated
material this correlation may be transmitted by d$^6$ electrons, if
the corresponding upper Hubbard band resides below the bottom of
the conduction band.

We thank J. Blinowski, P. Kacman, M. Kami\'nska, and A. Twardowski
for valuable discussions.  The work in Poland was supported by
State Committee for Scientific Research, Grant No. 2-P03B-02417,
and Foundation for Polish Science.


\begin{thebibliography}{99}
\bibitem{Remark} Because of compensation, the values of $x$
determined in Ref.~\cite{Sono01} from saturation values of
magnetization assuming $S=2$, are somewhat overestimated.
\end{thebibliography}
\end{document}